# A Novel Human Computer Interaction Platform based College Mathematical Education Methodology

Zhiyan Li
Department of Mathematics and Physics
Hohai University Changzhou Campus
Changzhou, P.R. of China

*Abstract*—This article proposes the analysis on novel human computer interaction (HCI) platform based college mathematical education methodology. Above for the application of virtual reality technology in teaching the problems in the study, only through the organization focus on the professional and technical personnel, and constantly improve researchers in development process of professional knowledge, close to the actual needs of the teaching can we achieve the satisfactory result. To obtain better education output, we combine the Kinect to form the HCI based teaching environment. We firstly review the latest HCI technique and principles of college math courses, then we introduce basic components of the Kinect including the gesture segmentation, systematic implementation and the primary characteristics of the platform. As the further step, we implement the system with the re-write of script code to build up the personalized HCI assisted education scenario. The verification and simulation proves the feasibility of our method.

## I. INTRODUCTION

Human-computer interaction research in the field of the computer science is a technical discipline, main information exchange and communication between human and computer. And in the field of the education teaching, human-computer interaction is regarded as the research and media interaction learning a technique or method of philosophy [1].

In field of computer science, human-computer interaction between human and computer through the exchange of mutual understanding and communication, and complete information management for people, with the maximum service and the processing, and other functions to make computer real people for studying the harmony of assistant on a technology science. In the field of education teaching, human-computer interaction can be simply defined as: in the process of teaching, education media and two-way information transmission between learners. This interaction is a teaching in the interaction between the main three factors, including between teachers and students, between teachers and media, students and media information, communication and feedback between dynamic processes. As the demonstration, we illustrate the application scenarios of the HCI technology in the figure one. Additionally, besides the pure naive HCI system, there are plenty of related techniques that will enhance the performance of the interaction feeling [2].

Virtual reality is of high and new technology in recent years, as also called artificial environment that is using the computer simulation to create a 3D virtual world, to provide users of the simulated senses such as vision, hearing, touch, let users as illustrates its general, timely, there is no limit to observe things in three dimensions [3-4].

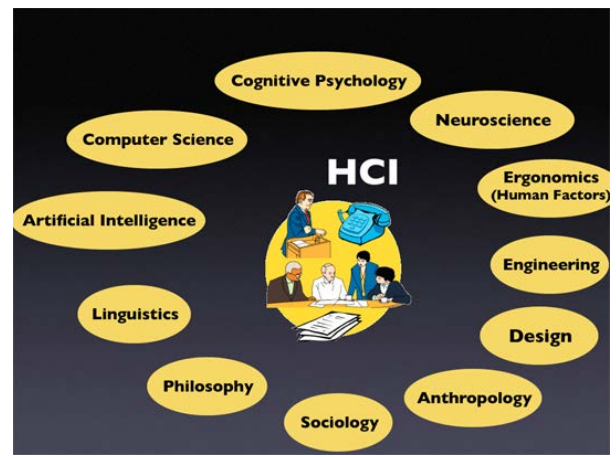

Fig. 1.  The Application Scenarios of the HCI Technology

However, there are still some challenges we should deal with before putting the technique to massive applications. (1) 3D data with high requirement: in order to make the virtual environment consistent with objective world, there are many different kinds of need of them the configuration of complex information to make accurate and complete description, there must be according to the requirement to collect teaching staff. At the same time, the need to study efficient modeling method, to rebuild its evolution and the various relationships between virtual objects and interaction, it also must have professional and technical development personnel to guide data collection according to the requirement of the modeling, the data is very large, it takes a lot of energy for data collection steps, sorting, organization, in order to get more realistic simulations [5]. (2) Virtual reality technology requirement is high. Virtual reality is a comprehensive integration technology, computer graphics, human-computer interaction technology, sensor technology, artificial intelligence, and other fields, technology integrated the technologies of the computer graphics, computer simulation technology, artificial intelligence, sensor technology, display technology, results of latest development of parallel processing technology such as network is a kind of technology aided by computer generated simulation system of the high and new technology [6-7]. (3) Computer hardware requirements. Virtual reality technology application of hardware conditions are too harsh, pattern formation is an important bottleneck in virtual reality, virtual reality is the most important feature of people can feel the scene control under vary the interaction of dynamic characteristics [8].

Above for the application of virtual reality technology in teaching the problems in the study, only through organization focus on professional and technical personnel, and constantly improve the researchers in development process of professional knowledge, close to the actual needs of the teaching, to collect accurate data in practice, strive to real simulation environment, and through the big scene model split, dynamic download and dynamic scheduling method, overcome bottleneck of computer hardware, developed in conformity with the actual needs of the teaching with virtual reality software. Therefore, in this paper, we conduct corresponding analysis on novel human computer interaction platform based college mathematical education methodology. The rest of the paper is organized as follows. In section 2, we review the state-of-the-art result on HCI and the corresponding technique background for the mathematical teaching paradigms. In the section 3, we propose the new way of combining HCI concepts with the education. In the section 4, we verify and implement the system with systematic analysis and simulation. In the section 5, we conclude the work with final summary and prospect.

## II. BACKGROUND ANALYSIS

### A. State-of-the-Art Review on HCI

During recent years, more and more related techniques are proposed for enriching overall HCI family. In [9], Jekaterina proposed the emotionally driven robot control architecture for the human-robot interaction. They enhance role of emotions in influencing human behavior in HRI as visual cues such as facial expressions are important in communication, research on emotion recognition, expression, and emotionally enriched the basic communication pattern. From the system as a whole, the information inside the machine processing and interactive process can be viewed as a process of inner loop their analysis achieved the better integration. In [10], Boskovic proposed the heuristic evaluation in the human computer interaction course. They pointed out that the evaluation supports other course objectives are combine the theoretical knowledge as usability principles to achieve better performance. In a loop, the person is the main body of information, responsible for information instruction as the circulation process of information acts as the control. In [11], the Fanello proposed the research on the depth camera for close-range human capture and interaction. They analyze the HCI system from the perspective of technique and the systematic design. More related research can be obtained in the literatures of [12-15]. In addition to basic HCI technique, the related research on the camera modelling [16-18], image processing [19-20] and data processing [21-25] should also be taken into consideration.

### B. The College Mathematical Education

Mathematics curriculum is the basic course of the higher education, the teaching quality directly affect the level of the school. Mathematics courses of study in addition to having a direct effect on future professional learning. More important, it can strengthen the students' rational thinking training, and improve students' logical thinking ability, strengthen students' ability of the computer, network communication. In fact, basic mathematics education will benefit the students for life.

Under this core background, we should enhance the current curriculum from the listed two perspectives. (1) The traditional academic teaching mode to play general college mathematics curriculum content is extensive and the comprehensive major advantages, it ignores the theory into practice and only pay attention to the importance of imparting theoretical knowledge as is hard to produce good teaching effect. Multidimensionality teaching mode and traditional teaching mode, the fundamental difference between is that it pays attention to both teaching and learning interaction, in the protection of teachers' emotional at the same time, and stimulate the students' learning enthusiasm and interest, as this teaching mode can improve the teaching effect of the college mathematics. (2) Multidimensionality, the implementation of the teaching mode can improve the teaching mode of college mathematics classroom is given priority to with teachers teach traditional ideas, fundamentally improve the past, monotonous teaching method, students in classroom order low participation related issues as able to adapt to the current curriculum under the basic requirement for teaching purpose. So, we must change the traditional teaching mode, to the multidimensional nature of the teachers and students in the teaching mode shift and this will become the focus of college mathematics teaching research work in the future.

## III. OUR METHODOLOGY

To finalize the proposed framework, we adopt the Kinect as the sensor for assisting achieving the HCI environment. In today's multimedia teaching, and many times the teacher need to constantly moving to the click of a mouse, keyboard for the computer under effective control, or need some of the handheld device. In many of the course at the same time, mainly take teachers as the main body, the students are passive listening, when the students just listen to below it with the teacher's interactive nature rarely. And with the emergence of device, we can use its motion-sensing technology, to obtain the depth information of human body, through the gesture recognition, to understand the operator's intentions, and to the operation of the computer effectively. And thus, this article is also based in complete device gesture recognition, on the basis of further realize natural user interface to build the body feeling teaching environment, improving the teaching of appeal.

Teaching design of the multimedia courseware is using the ideas and methods of the system theory, on the basis of the teaching goal, analysis the problems and basic requirements of teaching, to determine the steps to solve the problem, select the corresponding teaching strategies, including identifying the corresponding knowledge point sorting, select teaching media, design teaching environment, arrangement of teaching content and the present way to present information and basic feedback information, and human-computer interaction, etc. education information form of the organizational structure with the linear structure, branch structure and core network structure, etc. The traditional text, audio, video and other teaching materials of the information organization structure is linear typed. Multimedia courseware is a hypermedia structure, by similar to the human brain associative memory structure of the nonlinear network structure to organize education information that has no fixed order, also does not require people to extract the information in a certain order. Node, the chain, the network is the three basic elements of hypermedia structure. In the figure 2, we show the differences of GUI, virtual reality, ubiquitous computers and the augmented interaction.

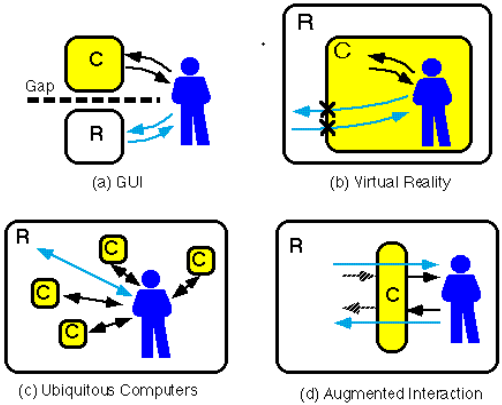

Fig. 2. The Differences of the GUI, VR, UC and AI

### A. The Features of Kinect

Motion-sensing technology is that the people can use body movements, directly interact with the peripheral device or the environment without using any complicated control equipment which can let people illustrates the situation and do interactive content. It is developed by Microsoft as a posture sensor input device, the structure of the Kinect is shown in the figure 3. It is mainly composed of a main photography head, a pair of depth sensors, a set of a microphone and a motor. Device with real-time dynamic capture, image recognition, microphone input, speech recognition, social interaction and other functions.

According to the device for the Windows SDK provides real-time 3D of the bone node location, it can obtain the node position at a certain moment, who told the angle between the node and the relative position. Continuous time case, the node can be obtained motion vector. Because mainly static posture recognition, so we can use point of view, and the node position to judge. The gesture recognition is divided into static gesture recognition and the dynamic gesture recognition. The primary implementation of the Kinect can be organized as the follows. (1) Speech recognition technology. Speech recognition is the voice as the research object, and let the machine by identifying and understanding the process of the speech signal into the corresponding text or command to make the person function naturally voice communication technology that belongs to the category of the multi-dimensional pattern recognition and with the intelligent computer interface. (2) The gesture recognition technology. Gesture recognition technology research mainly concentrated in the early to do a dedicated hardware device for input. Such as data glove, that is, people can wear on a similar gloves sensor, computer can get through it hand the rich information such as position and the primary condition of the extension of fingers.

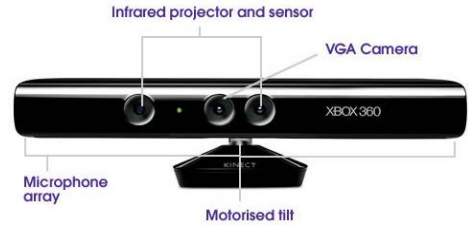

Fig. 3. The Components of the Kinect System

### B. Kinect HCI System Structure

Because it can use its depth cameras provide depth image, the pixels recorded the calibration of each point in the scene depth, resolution of a few centimeters. The depth of the camera can be very good to eliminate background noise, to extract the information. In the formula one, we show the hand gesture segmentation equation. Alpha is depth map of each pixel in the transparency, 255 said completely opaque, 0 means completely transparent. PlayerIndex denote as the character index number, DepthOfHand to hand the key points of the depth of the value, the depth of the depth of the corresponding pixel values in the depth map, the uLimit to limit depth value, dLimit depth for the lower limit value characteristics.

$$Alpha = 255 - \frac{255 \times (depth - DepthOfHand + dLimit)}{dLimit + uLimit} \quad (1)$$

After completion of gesture segmentation, we have taken pictures for the further processing. Algorithm steps set color threshold is first of all, when less than the threshold, the pixel is set to white. When is greater than the threshold, the pixel is set to the predefined a certain color. Then respectively from different directions, in turn, the scanning images of each pixel point, when the neighboring around two different values for pixels, the pixel as boundary point to find out the outline of the image. Dynamic hand gesture recognition method is based on the identification of the trajectory of the time series.

So in dynamic hand gestures recognition, the first and most important step in the hand from other image information extracted, remove the interference of background noise, the basic analysis to corresponding position, namely the gesture segmentation. It directly affects the results of the back of the dynamic gesture recognition accuracy, generally based on skin color segmentation, or wear a certain color of hand gloves on hand gesture segmentation in general. In this article, we mainly to gesture recognition is based on the device the SDK to get complete. It has provided the human bone three-dimensional location of each node, so the dynamic gesture training and the recognition need not extract the hand of information in the depth of each figure. But due to start in judging gestures is a combination of static gestures, namely to clench fist this action as a gesture of command. So still need their opponent gesture segmentation. The gesture segmentation strategy and the static gestures on the segmentation strategy is consistent, the first

PlayerIndex information provided by the device SDK, to pick up the man from the background. Secondly according to the depth limiting conditions will hand information extracted, the final opponent smooth processing of the information.

*C. The Script Implementation*

To implement the proposed approach, in the figure 4 we show the script implementation of the framework. The core part of the script is the event system. A node of the events of exports and other nodes between the entrance channel used to pass events called routing that can be bound by multiple nodes to form a system of events system is in addition to scene graph node hierarchy through another essential part of system events spread to get spread and cause the change of the node domain which affects the user's visual and cause the user to interact with the environment of interactive 3D experience. The figure 4 reflects part of the system and we will demonstrate the whole system architecture in the later research.

```
1  //Part of the Script for Our System
2  addChild(sp);
3  for (i=0; i<num+1; i++) {
4  for (j=0; j<num+1; j++) {
5  x0[i][j]=160*Math.sin(j*p/num)*Math.cos(i*2*p/num);
6  y0[i][j]=160*Math.cos(j*p/num);
7  z0[i][j]=160*Math.sin(j*p/num)*Math.sin(i*2*p/num);
8  } }
9  var map:Tuu=new Tuu(0,0);
10 var bmp:BitmapData=new BitmapData(map.width,map.height,false);
11 bmp.draw(map);
12 var anglex:Number=0,angley:Number=0.05;
13 stage.addEventListener(MouseEvent.MOUSE_DOWN,down);
14 var x1:Number=x0[i][j]*f/(f+z0[i][j]),
15 y1:Number=y0[i][j]*f/(f+z0[i][j]);
16 var x2:Number=x0[i][j+1]*f/(f+z0[i][j+1]),
17 y2:Number=y0[i][j+1]*f/(f+z0[i][j+1]);
18 var x3:Number=x0[i+1][j+1]*f/(f+z0[i+1][j+1]),
19 y3:Number=y0[i+1][j+1]*f/(f+z0[i+1][j+1]);
20 var x4:Number=x0[i+1][j]*f/(f+z0[i+1][j]),
21 y4:Number=y0[i+1][j]*f/(f+z0[i+1][j]);
22 var vArr:Vector.<Number>=
23 Vector.<Number>([x1,y1,x2,y2,x3,y3,x4,y4]);
24 var ind:Vector.<int>=Vector.<int>([0,1,3,1,2,3]);
```

Fig. 4. The Script Implementation of the Proposed System

IV. VERIFICATION AND IMPLEMENTATION

In this section, we verify the proposed system. The figure 5 illustrates the Kinect simulation on gesture extraction. The set of gesture recognition system based on it, in the teaching application, need to be defined in advance some basic gestures, and to add these gestures to training for training in the library, so that system can carry on the recognition of these gestures, then convert recognition of hand gestures to control command, instead of the keyboard. At the same time provided by the device's hand the node position of the mobile to control the movement of the mouse position, instead of mouse function. The effective combination of both has reached the non-contact control of the computer. Arrangement of the system in the classroom there are two ways, one kind of device is placed on the platform that is placed in the link to access the computer slightly above or below, so that teachers can by looking at the computer to operate, while operating the picture by projection projected onto a screen behind, because the regular teachers teaching are geared to the needs of students, so it is also more common. Correspondingly, we show the simulated HCI based math teaching condition in the figure 6.

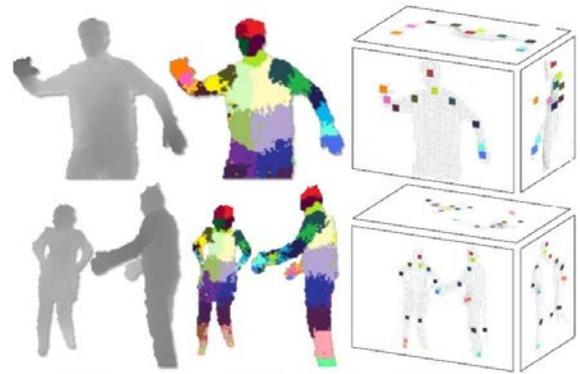

Fig. 5. The Kinect Simulation on Gesture Extraction

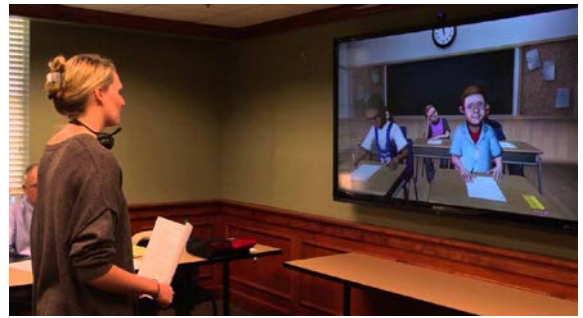

Fig. 6. The Simulated HCI based Math Teaching Condition

V. SUMMARY AND CONCLUSION

In this paper, we conduct basic analysis on the novel human computer interaction platform based college mathematical education methodology. This interaction is a teaching in the interaction between the main three factors, including between teachers and students, between teachers and media, students and media information, communication and feedback between dynamic processes. We propose the Kinect based HCI system under basis of the state-of-the-art reviews of other approaches. And thus, this article is also based in complete device gesture recognition, on the basis of further realize natural user interface to build the body feeling teaching environment, improving the teaching of appeal. We implement and re-write the core part of the script to achieve the task of the interaction teaching. The verification part reflects the effectiveness of our system. In the later research, we will focus on the enhancement of interaction efficiency for overall optimization.